\documentclass[usegraphicx,usenatbib]{mn2e}
\begin{document}

\title[Testing thermal reprocessing in AGN accretion discs]{Testing thermal reprocessing in AGN accretion discs}

\author[Cackett, Horne \& Winkler]
{
Edward M. Cackett$^{1,2}$\thanks{ecackett@umich.edu}, Keith Horne$^2$ and Hartmut Winkler$^3$
\\ $^1$ Dean McLaughlin Postdoctoral Fellow, Department of Astronomy, University of Michigan, Ann Arbor, MI 48109, USA
\\ $^2$ SUPA, School of Physics and Astronomy,
	University of St.~Andrews,
	KY16 9SS, Scotland, UK
\\ $^3$ Department of Physics, University of Johannesburg, PO Box 524, 2006 Auckland Park, 
	South Africa
}
\date{Accepted . Received ; 
in original form }

\maketitle

\begin{abstract}
The thermal reprocessing hypothesis in AGN, where EUV/X-ray photons are reprocessed by the accretion disc into optical/UV photons, predicts wavelength-dependent time delays between the optical continuum at different wavelengths.  Recent photometric monitoring by Sergeev et al. has shown that the time-delay is observed in 14 AGN, and generally seen to increase with increasing wavelength, as predicted in the reprocessing scenario.  We fit the observed time delays and optical spectral energy distribution using a disc reprocessing model.  The model delivers estimates for the nuclear reddening, the product of black hole mass times accretion rate, and the distance to each object.  However, the distances at face value give H$_0 = 44 \pm 5$  km s$^{-1}$ Mpc$^{-1}$ - a factor of 1.6 smaller than generally accepted.  We discuss the implications of this on the reprocessing model. 

\end{abstract}

\begin{keywords}
galaxies: active -- 
galaxies: nuclei -- galaxies: Seyfert
\end{keywords}

\section{Introduction}

Echo mapping, or reverberation mapping, \citep{blandmckee82,peterson93} uses light travel time to measure spatial separations within a distant accretion flow.  Ionizing photons produced close to the compact object irradiate the surrounding gas which reprocesses these into UV/optical continuum and emission lines.  As the ionizing radiation varies erratically, so do the reprocessed components but with time delays due to light travel time within the system - the light travel time from source to reprocessing site to observer is longer than that on the direct path from source to observer.  These observable delays provide indirect information on the size and structure of the surrounding accretion flows.  The micro-arcsecond resolution that can be achieved by echo mapping is unreachable by any other technique (1 light day across at 100 Mpc is $\sim$2 micro-arcsecond).

It has long been known that the optical and UV continuum in AGN are highly correlated and vary nearly simultaneously.  Initial attempts at measuring the time-delay (or lag) between the optical and UV continuum found it to be less than a couple of days \citep[e.g.,][]{stirpe94}.  However, higher time sampling by later spectroscopic monitoring campaigns have allowed for the measurement of wavelength-dependent continuum time delays in two objects (NGC~7469 and Ark~564).  These two objects appear to follow the predicted relation for an irradiated accretion disc \citep{wandersetal97,collier98,collier01}, with the lags being of the order of 1-2 days.  Such small lags can only be explained in terms of AGN reprocessing as disc instabilities should produce time delays of the order of the viscous timescale (the timescale on which the local surface density changes) which is at least hundreds of days in AGN for the region where the optical emission peaks \citep{kroliketal91}.

Recent photometric monitoring by \citet{sergeev05} showed significant lags between optical lightcurves in 14 AGN, with the lags between the lightcurves determined by standard cross-correlation techniques.  Their measured lags range from tenths of a day to several days. The lag increases with wavelength, as is predicted by reprocessing in an accretion disc, where the inner hotter regions see the ionizing source before the outer cooler regions.  They also find that the delay is systematically greater for greater absolute nuclear luminosity, following $\tau \propto L^b$ where $b \approx 0.4-0.5$, consistent with a disc reprocessing model.  The lags are interpreted in terms of the light-travel time from an ionizing source above the disc to the region in the disc where the ionizing radiation is reprocessed into optical continuum emission.  Thus, the brighter the source, the greater the light-travel time to the region emitting the optical continuum.

\citet{collier99} showed how measuring time delays between the optical/UV continuum in an AGN accretion disc and the flux from the disc, the distance to an AGN can be measured, and hence AGN can be used as standard candles to determine the current expansion rate of the Universe, Hubble's constant, $H_0$.  From measurements of the wavelength-dependent time delays in NGC~7469, these authors determine $H_0 = 42 \pm 9$ km s$^{-1}$ Mpc$^{-1}$ assuming an inclination of $45^{\circ}$.  However, this is at odds with other determinations of $H_0$.  Over the past decade, several methods have been employed to determine $H_0$, with all methods now converging to a value of around 72 km s$^{-1}$ Mpc$^{-1}$ \citep[e.g.][]{freedman01}.  The method using AGN accretion discs does not rely on the distance ladder calibration, and thus provides another independent check on these results.  One distinct advantage of this method, compared to other methods, is that AGN are common in the Universe, and can be found over a wide range of redshifts.  As they are some of the most energetic sources in the Universe, AGN can be seen at redshifts beyond the supernova horizon, and can thus potentially set tight constraints on $\Omega_M$ and $\Omega_\Lambda$ -- it is at redshifts greater than $z \sim 1$ where large differences are seen between different cosmological models.

In this paper, we re-analyse the \citet{sergeev05} lightcurves, also in terms of thermal reprocessing in an AGN disc, in an attempt to extend the method of \citet{collier99} to use AGN accretion discs as a cosmological probe.  In Section~2 we discuss the AGN lightcurves that we analyse and in Section~3 we detail the accretion disc model used.  The results of our model fits to the data are presented in Section~4, and the implications of these results discussed in Section~5.

\section{AGN lightcurves}

We use the published optical lightcurves for 14 AGN from \citet{sergeev05} in the B, V, R, R1 and I bands.  The filters used are non-standard filters \citep{doroshenko05a,doroshenko05b}, though their B, V, and R are close to Johnson filters, the R1 is close to the Cousins I filter and the I filter is close to the Johnson I filter.  The published lightcurves are given as relative fluxes with respective to comparison stars.  However, the magnitudes of the comparison stars are only given in the standard B, V, R, I bands, rather than the non-standard bandpasses used.  For this work, we require the absolute fluxes of the AGN, rather than the relative fluxes, and therefore require the magnitudes of the comparison stars in the filters used.  To calculate these we use {\sc Xcal}, Keith Horne's interactive synthetic photometry program.  {\sc Xcal} computes predicted count rates from target spectra in standard or non-standard passbands, allowing stellar spectra to be fit to the comparison star magnitudes in the standard Johnson-Cousins filters.  Spectra of main sequence stars from the Bruzual-Persson-Gunn-Stryker atlas \citep{gunnstryker83} were individually scaled to fit the observed magnitudes, and the star with lowest $\chi^2$ was adopted to model the spectrum of the comparison star.  After finding the best-fit stellar spectrum, an intergral over the non-standard passband gives the magnitudes.  In this way, we determined the fluxes of the comparison stars in the B, V, R, R1 and I filters used by \citet{sergeev05} (see Table \ref{tab:comp_stars} and thereby convert the AGN lightcurves from differential magnitudes to absolute fluxes.
\begin{table}
\centering
\caption{Redshifts, $z$, and fluxes (in mJy) for the comparison stars used by \citet{sergeev05} for each of the 14 AGN in their B, V, R, R1 and I filters.}
\label{tab:comp_stars}
  \begin{tabular}{ccccccc}
    \hline
	Object & z & B & V & R & R1 & I\\
    \hline
NGC 4051    & 0.002   & 5.1  & 7.9  & 9.3  & 9.7   & 9.8 \\
NGC 4151    & 0.003   & 8.7  & 26.6 & 50.8 & 66.4  & 70.5 \\
NGC 3227    & 0.004   & 15.2 & 28.5 & 36.7 & 39.5  & 40.1 \\
NGC 3516    & 0.009   & 10.1 & 15.6 & 18.7 & 19.4  & 19.4 \\
NGC 7469    & 0.016   & 26.5 & 55.2 & 77.2 & 88.0  & 90.2 \\
NGC 5548    & 0.017   & 6.0  & 11.1 & 14.4 & 15.5  & 15.7 \\
Mrk 6       & 0.019   & 3.5  & 5.9  & 7.5  &  8.0  & 8.1 \\
MCG+8-11-11 & 0.020   & 6.3  & 19.2 & 36.7 & 48.0  & 50.9 \\
Mrk 79      & 0.022   & 22.0 & 37.6 & 47.9 & 51.0  & 51.3 \\
Mrk 335     & 0.026   & 8.0  & 13.6 & 16.9 & 18.1  & 18.3 \\
Ark 120     & 0.032   & 4.5  & 10.2 & 15.0 & 17.1  & 17.5 \\
Mrk 509     & 0.034   & 25.0 & 40.7 & 49.2 & 51.4  & 51.8 \\
3C 390.3    & 0.056   & 3.8  & 7.0  & 9.0  & 9.9   & 10.0 \\     
1E 0754.6+3928 & 0.096 & 5.8  & 10.9 & 14.0 & 15.1  & 15.3 \\

    \hline
  \end{tabular}
\end{table}

\begin{figure*}
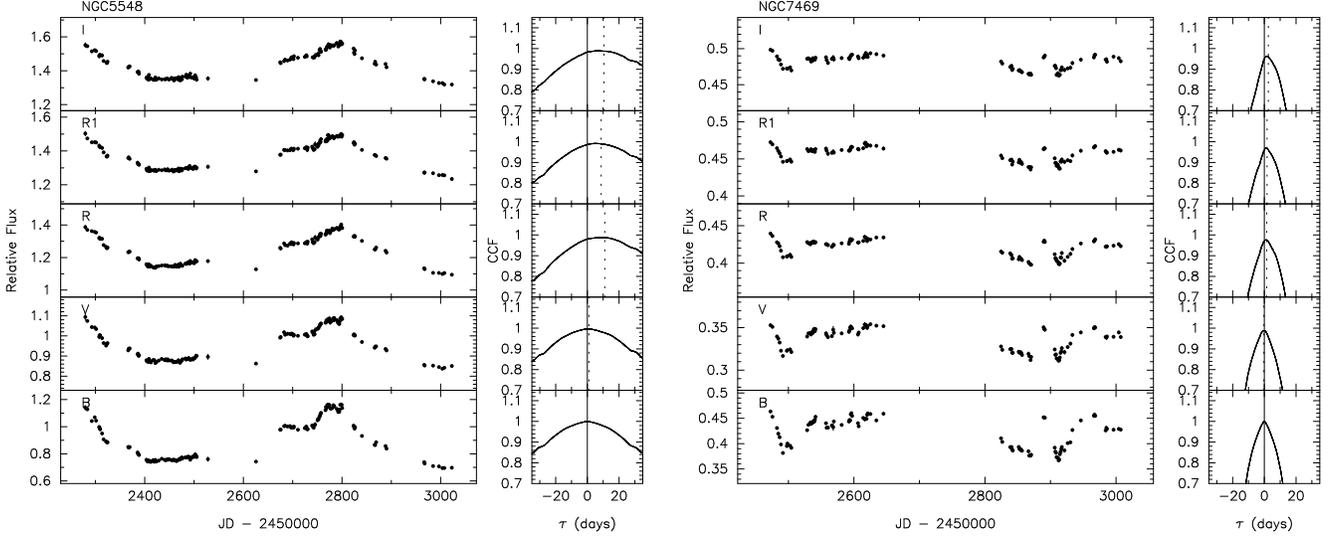

 \centering
 \includegraphics[width=85mm]{ngc5548_lc_individual.ps}
 \hspace{3mm}
 \includegraphics[width=85mm]{ngc7469_lc_individual.ps}
 \caption{Lightcurves for NGC 5548 and NGC 7469 in all 5 bands. Corresponding cross-correlation functions (CCF) for the lightcurves with respect to the B-band are also shown.  In the B-band the auto-correlation function is plotted.  Solid line indicates a time-delay, $\tau$ = 0.0 d, and the dotted line shows the centroid of the CCF.}
 \label{fig:individ_lc}
\end{figure*}

All 14 AGN show significant variability over the approximately 3 year monitoring period \citep[see fig.~1 from][]{sergeev05}.  The median temporal sampling for the lightcurves is 2 to 3 days \citep[see table~3 from][for sampling rates for each AGN]{sergeev05}.  As an example of the strong correlations between the lightcurves in the different wavebands, all 5 lightcurves for NGC 5548 and NGC 7469, with the cross-correlation functions, are shown in Fig. \ref{fig:individ_lc}.  We have determined the lags between the lightcurves (relative to the B-band lightcurve) for each AGN using the interpolation cross-correlation method \citep{gaskellpeterson87,white94}, and find the delays to be the same as determined by \citet{sergeev05} (see their Table 4).  The peak time delay is generally smaller than the centroid time delay, though in many cases they are consistent within the uncertainties.  The maximum and minimum fluxes of each of the lightcurves are also given in Table \ref{tab:maxminflux} as these are used in the model which we describe later.

\begin{table}
\centering
\caption{Maximum and minimum fluxes (mJy) from the AGN lightcurves.  Continued on next page.}
\label{tab:maxminflux}
\begin{tabular}{lccc}
    \hline
	Object & Filter & Maximum flux & Minimum flux \\
	  & & (mJy) & (mJy) \\
    \hline
NGC 4051 & B & $12.89 \pm 0.12$ & $9.92 \pm  0.05$ \\
 & V & $21.46 \pm 0.16$ & $18.22 \pm 0.08$ \\
 & R & $34.23 \pm 0.19$ & $29.71 \pm 0.16$ \\
 & R1 & $41.31 \pm 0.27$ & $36.55 \pm 0.20$ \\
 & I & $43.42 \pm 0.30$ & $38.94 \pm 0.24$ \\
NGC 4151 & B & $57.53 \pm 0.54$ & $29.93 \pm 0.29$ \\
 & V & $74.97 \pm 0.35$ & $49.07 \pm  0.27$ \\
 & R & $117.93 \pm 0.49$ & $84.37 \pm 0.34$ \\
 & R1 & $125.98 \pm 0.54$ & $91.60 \pm 0.38$ \\
 & I & $133.10 \pm 0.56$ & $97.44 \pm  0.43$ \\
NGC 3227 & B & $11.41 \pm 0.07$ & $8.69 \pm 0.05$ \\
 & V & $22.84 \pm 0.13$ & $19.53 \pm 0.10$ \\
 & R & $41.02 \pm 0.20$ & $36.95 \pm 0.18$ \\
 & R1 & $51.03 \pm 0.26$ & $46.72 \pm  0.23$ \\
 & I & $54.54 \pm 0.27$ & $50.11 \pm 0.52$ \\
NGC 3516 & B & $15.85 \pm 0.10$ & $12.03 \pm 0.06$ \\
 & V & $31.00 \pm 0.13$ &  $27.33 \pm  0.13$ \\
 & R & $52.15 \pm 0.22$ & $46.25 \pm 0.20$ \\
 & R1 & $62.06 \pm 0.39$ & $56.58 \pm 0.46$ \\
 & I & $65.25 \pm 0.36$ & $59.74 \pm 0.30$ \\
NGC 7469 & B & $12.28 \pm 0.06$ & $9.72 \pm 0.04$ \\
 & V & $19.53 \pm 0.09$ & $17.16 \pm 0.07$ \\
 & R & $33.95 \pm 0.14$ & $30.73 \pm 0.13$ \\
 & R1 & $41.56 \pm 0.17$ & $38.33 \pm 0.16$ \\
 & I & $44.99 \pm 0.18$ & $41.76 \pm 0.17$ \\
NGC 5548 & B & $6.94 \pm  0.05$ & $4.15 \pm  0.02$ \\
 & V & $12.20 \pm  0.19$ & $9.33 \pm  0.04$ \\
 & R & $20.17 \pm  0.11$ & $15.73 \pm  0.09$ \\
 & R1 & $23.24 \pm  0.21$ & $19.09 \pm  0.13$ \\
 & I & $24.71 \pm  0.10$ & $20.71 \pm  0.15$ \\
Mrk 6 & B &  $4.50 \pm 0.03$ & $2.99 \pm 0.02$ \\
 & V & $9.42 \pm 0.06$ & $7.25 \pm 0.04$ \\
 & R & $17.05 \pm 0.09$ & $13.45 \pm 0.06$ \\
 & R1 & $19.07 \pm 0.11$ & $15.50 \pm 0.08$ \\
 & I & $20.26 \pm 0.10$ & $16.44 \pm 0.09$ \\
MCG+8-11-11 & B & $6.83 \pm 0.04$ & $3.61 \pm 0.03$ \\
& V & $10.60 \pm 0.05$ & $7.08 \pm 0.05$ \\
& R & $19.61 \pm 0.08$ & $14.98 \pm 0.07$ \\
& R1 & $22.51 \pm 0.09$ & $17.53 \pm 0.08$ \\
& I & $24.44 \pm 0.10$ & $19.16 \pm 0.09$ \\
Mrk 79 & B & $6.74 \pm 0.05$ &  $4.02 \pm 0.02$ \\
 & V & $9.65 \pm  0.08$ &  $6.87 \pm 0.04$ \\
 & R & $15.12 \pm 0.09$ & $11.50 \pm 0.05$ \\
 & R1 & $16.22 \pm 0.12$ & $12.67 \pm 0.06$ \\
 & I & $16.96 \pm 0.13$ & $13.32 \pm 0.08$ \\
Mrk 335 & B & $7.61 \pm 0.06$ & $4.45 \pm 0.02$ \\
& V & $8.90 \pm 0.06$ & $6.04 \pm 0.04$ \\
& R & $12.67 \pm 0.07$ & $9.00 \pm 0.04$ \\
& R1 & $12.09 \pm 0.07$ & $8.74 \pm 0.04$ \\
& I & $12.58 \pm 0.08$ & $9.10 \pm 0.05$ \\
Akn 120 &  B & $12.55 \pm 0.13$ & $7.032 \pm 0.10$ \\
 & V & $16.37 \pm  0.13$ & $10.60 \pm 0.08$ \\
 & R & $23.55 \pm 0.13$ & $16.00 \pm  0.09$ \\
 & R1 & $24.02 \pm 0.15$  & $17.04 \pm 0.09$ \\
 & I & $25.02 \pm 0.16$ & $18.06 \pm 0.10$ \\
Mrk 509 & B & $15.79 \pm 0.11$ & $12.11 \pm 0.09$ \\
 & V & $18.07 \pm 0.11$ & $15.02 \pm  0.11$ \\
 & R & $25.40 \pm 0.12$ & $22.24 \pm 0.11$ \\
 & R1 & $23.57 \pm 0.13$ & $20.76 \pm 0.12$ \\
 & I & $24.47 \pm 0.14$ & $21.71 \pm 0.12$ \\

\hline
\end{tabular}
\end{table}

\setcounter{table}{1}
\begin{table}
\centering
\caption{continued.
\hfil}
\begin{tabular}{lccc}
    \hline
	Object & Filter & Maximum flux & Minimum flux \\
	  & & (mJy) & (mJy) \\
    \hline
3C 390.3 & B & $2.86 \pm 0.02$ & $1.05 \pm 0.01$ \\
 & V & $4.12 \pm 0.02$ & $2.19 \pm 0.01$ \\
 & R & $6.30 \pm 0.03$ & $3.74 \pm 0.02$ \\
 & R1 & $6.77 \pm 0.04$ & $4.25 \pm 0.03$ \\
 & I & $7.13 \pm 0.05$ & $4.53 \pm 0.03$ \\
1E 0754.6+3928 &  B &  $5.12 \pm 0.03$ & $4.45 \pm 0.03$ \\
  & V &  $6.87 \pm 0.04$ & $6.33 \pm 0.06$ \\
  & R &  $8.33 \pm  0.05$ & $7.62 \pm  0.04$ \\
  & R1 &  $9.40 \pm 0.06$ & $8.58 \pm  0.05$ \\
  & I &  $8.81 \pm 0.05$ &  $8.11 \pm  0.06$ \\

\hline
\end{tabular}
\end{table}

\section{Accretion disc model}

\citet{sergeev05} successfully modelled the lag-luminosity relation in these sources with a disc reprocessing model, where interband time delays exist due to different light travel times from the ionizing source and the different continuum emitting regions.  Here we take this further, by simultaneously fitting the interband lag distribution and spectral energy distribution (SED) in terms of a disc reprocessing model.  In this model we fit the SED at the maximum and minimum points in the lightcurve, interpreting the change in brightness as a change in the luminosity of the ionizing source driving the reprocessing in the disc, and hence a change in disc temperature.  Simultaneously we fit the lags with the same model.

We assume that there is a highly variable source of ionizing radiation near the disc axis that illuminates the disc, driving optical/UV variability and leading to wavelength-dependent time delays.  The hot inner regions which emit mainly UV photons `see' the driving ionizing photons before the cool outer regions which emit mainly optical photons.  Thermal radiation from a disc annulus at temperature $T(R)$ emerges with a range of wavelengths, centred at $\lambda \sim hc/kT(R)$. Roughly speaking, each wavelength picks out a different temperature zone and the time delay $\tau = R/c$ measures the corresponding radius. Thus, shorter wavelengths sense disc annuli at higher temperatures.

In this model there is heating due to irradiation as well as viscous heating in the disc.  For viscous heating alone, when the radius is much greater than the inner most stable orbit, the disc temperature profile should follow \begin{equation}
T(R) = \left(\frac{3GM\dot{M}}{8\pi R^3 \sigma}\right)^{1/4}
\end{equation}
where $M$ is the mass of the black hole, and $\dot{M}$ the mass accretion rate.  In the case of irradiation of the disc from a height $H_x$ above the disc, the temperature profile is a combination of the temperature due to irradiation and the temperature due to viscous heating,
\begin{equation}
T(R) = \left[\left(\frac{3GM\dot{M}}{8\pi R^3 \sigma}\right) + \left(\frac{(1-A)L_x}{4\pi \sigma R_\ast^2}\right)\cos\theta_x \right]^{1/4}\; ,
\end{equation}
where $L_x$ is the luminosity of the irradiating source, $A$ is the disc albedo, $R_\ast$ is the distance from the illuminating source to the surface element of the disc, and $\theta_x$ is the angle between the disc surface normal and a unit vector from the surface element to the variable central source.  This leads to $T(R) \propto (L_x H_x)^{1/4}R^{-3/4}$, for $R \gg H_x$ and provided the disc thickness is $H \ll H_x$ \citep*[e.g.][]{frankkingraine}.  Thus, in both cases, we expect $T(R) \propto R^{-3/4}$.  As the time delay goes as $\tau = R/c$ and the wavelength goes like $\lambda \propto T^{-1}$, then this predicts that the wavelength-dependent time-delay profile follows a $\tau \propto \lambda^{4/3}$ relationship.

With the temperature profile determined from the wavelength-dependent time delays, the predicted spectrum can be determined by summing up blackbodies over disc annuli, and comparing this with the observed spectrum leads to a determination of the distance to the source \citep{collier99}.  In this way the predicted spectrum is given by
\begin{equation}
f_\nu = \int_{R_{in}}^{R_{out}} B_{\nu} \; {\mathrm d}\Omega = \int_{R_{in}}^{R_{out}} \frac{2hc}{\lambda^3}\frac{1}{e^{hc/\lambda kT} - 1}\frac{2\pi R \;{\mathrm d}R \cos i}{D^2}
\label{eq:BB}
\end{equation}
with $B_\nu$ the Planck function, ${\mathrm d}\Omega$ the solid angle subtended by the disc annuli, $R_{in}$ and $R_{out}$ the inner and outer disc radii, $i$ the inclination angle of the disc, and $D$ the distance.  Following \citet{collier_thesis}, we assume a general temperature profile for the disc of $T = T_0 y^{-b}$ with $y = R/R_0$ ($T_0$ is the temperature at radius $R_0$), and $s_0 = hc/\lambda k T_0$ such that $s = s_0 y^b$ and ${\mathrm d}y = y \; {\mathrm d}s/bs$.  Over a range of wavelengths that are well within those corresponding to the inner and outer disc radii, substituting these expressions into Eq. \ref{eq:BB} leads to
\begin{equation}
f_\nu = \frac{4\pi R_0^2 \cos i}{D^2} (hc)^{1 - (2/b)} \lambda^{(2/b) - 3} (kT_0)^{2/b} \frac{I_2(b)}{b} \; \; ,
\label{eq:fnu}
\end{equation}
where
\begin{equation}
I_n(b) = \int_0^\infty \frac{s^{(n/b) - 1}\;{\mathrm d}s}{e^s - 1} \: .
\end{equation}
$I_n(b)$ is a numerical function, and we calculate $I_2(3/4) = 1.93$ and $I_3(3/4) = 6.49$ ($b=3/4$ corresponds to the standard thin disc assumption). 

AGN spectra, however, rarely match the classical thin-disc spectrum $f_\nu \propto \nu^{1/3}$.  This difference is likely due to contributions from stars and dust.  Nevertheless, the disc spectrum can be isolated by taking difference spectra, and \citet{collier99} found that the \textit{variable} component of AGN light in NGC~7469 does have $\Delta f_\nu \propto \nu^{1/3}$.  For this reason we choose to model the difference spectra of the AGN using the method described below.

\subsection{Disc Transfer Function}
\begin{figure}
 \centering
 \includegraphics[width=8cm]{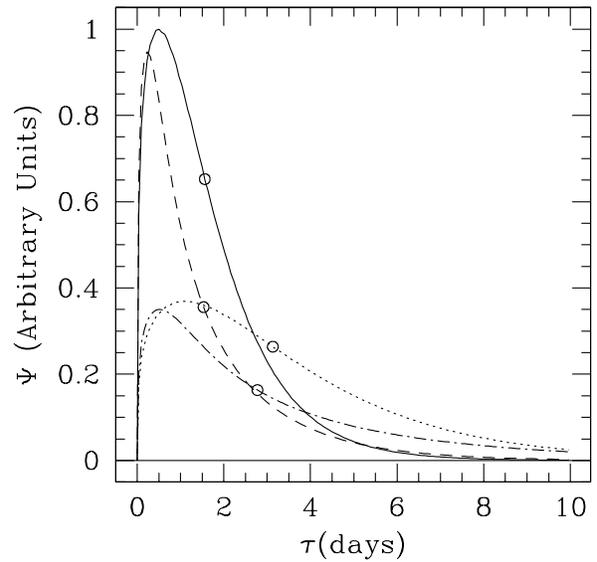}
 \caption{Irradiated accretion disc transfer functions for 4400${\rm \AA}$ with disc inclinations of $0^{\circ}$ (solid) and $45^{\circ}$ (dashed) and 8000${\rm \AA}$ with disc inclinations of $0^{\circ}$ (dotted) and $45^{\circ}$ (dash-dot).  $T_B$ = 15000 K and $T_F$ = 12000 K.  Circles mark the centroid delay for comparison with the peaks.}
 \label{fig:disctransfunc}
\end{figure}

We model the AGN wavelength-dependent time-delay distribution and spectrum using a disc transfer function $\Psi_\nu(\tau,\lambda)$ to describe the response of the disc to changes at a given wavelength and time delay, as outlined by \citet{collier_thesis}.  The disc transfer function is defined as
\begin{eqnarray}
\Psi_\nu(\tau,\lambda) &=& \int^{R_{out}}_{R_{in}}\frac{\partial B_\nu}{\partial T} \frac{\partial T}{\partial L_x} \times \nonumber \\
& & \delta(\tau - \frac{R}{c}(1+ \sin i \cos \theta)) \; {\mathrm d}\Omega\\
&=& \int^{R_{out}}_{R_{in}} \int^{2\pi}_{0} \frac{\partial B_\nu}{\partial T} \frac{\partial T}{\partial L_x} \frac{R \; {\mathrm d}R \; {\mathrm d}\theta \cos i}{D^2} \times \nonumber \\
& &\delta (\tau - \frac{R}{c}(1+ \sin i \cos \theta)) \; ,
\end{eqnarray}
where $L_x$ is the driving ionizing luminosity.  The $\delta$-function ensures that only radii corresponding to the specific time delay contribute to the transfer function.  As we are fitting the maximum and minimum SED of the AGN with this model, we treat the disc as having temperature $T_B$ at $R_0$ during the bright state and $T_F$ at $R_0$ during the faint state.  Hence, 
\begin{equation}
\frac{\partial B_\nu}{\partial T} \frac{\partial T}{\partial L_x} = B_\nu(T_2) - B_\nu(T_1)
\end{equation}
and $T_2 = T_B(R/R_0)^{-b}$, $T_1 = T_F(R/R_0)^{-b}$, and we take $R_0 = 1$ light-day.  Examples of the transfer function are given in Fig. \ref{fig:disctransfunc}.

\subsection{Time delays}
The centroid of the transfer function is equivalent to the centroid of the CCF which is used as a measure of the luminosity-weighted radius of the reprocessing region \citep[]{koratkar91}, however, see \citet{robinsonperez90} and \citet{welsh99} for limitations. For our blackbody disk model, the time delay centroid of the transfer function is:
\begin{equation}
\langle \tau \rangle = \frac{\int \tau \Psi_\nu(\tau,\lambda)\;{\mathrm d}\tau}{\int \Psi_\nu(\tau,\lambda)\;{\mathrm d}\tau} \; ,
\end{equation}
which becomes
\begin{equation}
\langle \tau \rangle = \frac{\int^{R_{out}}_{R_{in}} [B_\nu(T_2) - B_\nu(T_1)](R^2/c) \; {\mathrm d}R}{\int^{R_{out}}_{R_{in}}[B_\nu(T_2) - B_\nu(T_1)]R \; {\mathrm d}R} \; .
\end{equation}
This reduces to
\begin{equation}
\langle \tau \rangle = \frac{R_0}{c}\left(\frac{\lambda k T_B}{hc} \right)^{1/b} \frac{I_3(b)}{I_2(b)}\frac{1 - \varepsilon^{3/2}}{1 - \varepsilon} \; ,
\label{eq:taudelay}
\end{equation}
where $\varepsilon = (T_F/T_B)^{2/b}$.  Rearranging the above equation for $kT_B$ and substituting into Eq. \ref{eq:fnu}, the bright state disc spectrum is given in terms of the measured lag:
\begin{equation}
f_\nu^B = \frac{4\pi \cos i}{D^2} hc^3 \lambda^{-3} \langle\tau\rangle^2
  \frac{I_2(b)^3}{b I_3(b)^2}\left(\frac{1 - \varepsilon}{1 - \varepsilon^{3/2}}  \right)^2 \; .
\label{eq:flux_B}
\end{equation}
Similarly, it can be shown that the faint state spectrum is just given by $f_\nu^{F} = \varepsilon f_\nu^B$ and the difference spectrum between the bright and faint state is given by $\Delta f_\nu = f_\nu^B - f_\nu^F = (1 - \varepsilon) f_\nu^B$.  As here we will always use $b = 3/4$ (as is appropriate for a classical thin accretion disc), we evaluate Eq. \ref{eq:taudelay} and \ref{eq:flux_B} for $b = 3/4$ below:
\begin{equation}
\frac{\langle \tau \rangle}{0.62~ \mathrm{d}} = \left(\frac{\lambda}{10^4 {\mathrm \AA}}\right)^{4/3} \left(\frac{T_B}{10^4 {\mathrm K}} \right)^{4/3} \frac{1 - \varepsilon^{3/2}}{1 - \varepsilon} \; ,
\end{equation}

\begin{eqnarray}
\frac{f_\nu^B}{40.1 \; \mathrm{Jy}} &=&  \left(\frac{\tau}{\mathrm{d}}\right)^2 \left(\frac{\lambda}{10^4 {\mathrm \AA}}\right)^{-3}\left(\frac{D}{\mathrm{Mpc}}\right)^{-2}  \cos i \times \nonumber \\
& & \left(\frac{1 - \varepsilon}{1 - \varepsilon^{3/2}}  \right)^2 \; .
\end{eqnarray}
Therefore, if the bright and faint state fluxes and interband time delays are measured from monitoring observations of an AGN, then the distance to the AGN can be determined:
\begin{eqnarray}
\frac{D}{6.3 \mathrm{Mpc}} &=&  \left(\frac{\tau}{\mathrm{d}}\right) \left(\frac{\lambda}{10^4 {\mathrm \AA}}\right)^{-3/2} \left(\frac{f_\nu^B/\cos i}{\mathrm{Jy}}\right)^{-1/2} \times \nonumber \\ & &
\left(\frac{1 - \varepsilon}{1 - \varepsilon^{3/2}}  \right)^2 \; .
\end{eqnarray}
Thus from the reverberation results we obtain a distance estimate
that is independent of reliance on the Hubble law, and so can be used to
test cosmological predictions for the luminosity distance versus redshift
relationship (see Section 4.1 and 5).  We also note that in the above equation all factors of $(1+z)$ to change from observed to emitted quantities cancel out.

\subsection{Applying this model to the Sergeev et al. data}

With interband time delays and long-term monitoring lightcurves for 14 AGN, the \citet{sergeev05} data offer an opportunity to test this reprocessing model and to determine distances to a sample of 14 AGN.  We fit the difference spectra (between the maximum and minimum points on the lightcurves) with the above accretion disc model.  An additional component to the model not discussed above is extinction due to interstellar reddening within our own Galaxy and also the within the AGN - this is required to correct the difference spectra to show a $\nu^{1/3}$ slope.  Therefore, as part of the model, we include $E(B-V)$ as a free parameter (we discuss this assumption below). The other free parameters in the fit to the data are the bright state temperature, $T_B$, the faint state temperature, $T_F$, the distance to the AGN, $D$, and the host galaxy fluxes in the V, R, R1 and I bands.  For a lower limit on the distance (and hence an upper limit on $H_0$), we set the host galaxy contribution in the B band to zero.  Whilst this is clearly an incorrect assumption, without it, a fit would be degenerate since a fainter accretion disc can be offset by a brighter galaxy. Independent measurements of the host galaxy in the bands, would constrain these nuisance parameters to reduce or remove this degeneracy.  In the model we assume that the accretion disc inclination angle, $i = 45^\circ$, as would be expected for type I AGN, where the broad emission lines are visible, from unified AGN theories \citep[e.g.,][]{antonucci93}.  However, the affect on the distance determined is small as $D \sim (\cos i)^{1/2}$.

\subsection{Extinction within the central regions of AGN}

\begin{figure*}
\centering
\includegraphics[width=17cm]{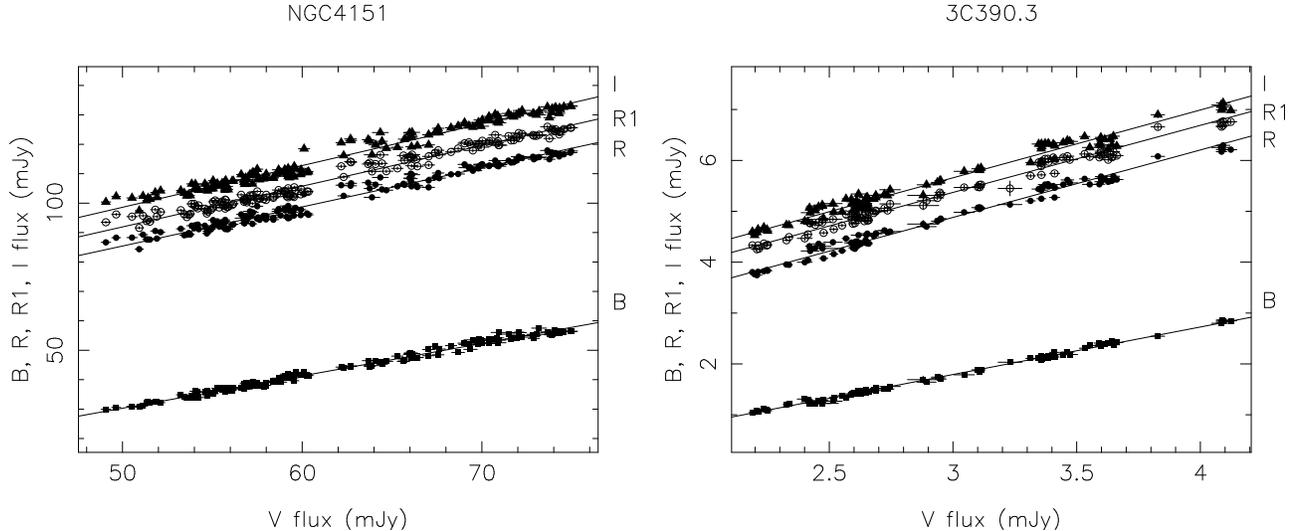}
\caption{Flux-flux diagrams for NGC 4151 (left) and 3C~390.3 (right).  B (squares), R (filled circles), R1 (open circles) and I (triangles) flux is plotted against V flux at corresponding times.  The best straight line fits are also shown.  NGC 4151 and 3C~390.3 are shown as these objects display the largest variability.}
\label{fig:fluxflux}
\end{figure*}

Two of the assumptions made when applying this model are that the discrepancy in slope between the difference spectra and a $\nu^{1/3}$ slope is due to extinction from our own Galaxy and within the central regions of the AGN, and also that the variable component of the AGN spectra does not change shape between the bright and the faint state.  To test these assumptions, we use the flux variation gradient method of \citet{winkleretal92} and \citet{winkler97}.  In this method, when comparing the flux through one filter with the flux through another filter at corresponding times, a very strong linear relationship is seen. We show a couple of examples of `flux-flux' diagrams in Fig. \ref{fig:fluxflux} obtained with the Sergeev et al. data.  We choose to show flux-flux diagrams for the two most variable AGN in the sample (3C~390.3 and NGC 4151).  Firstly, as the relationship is so close to linear, it demonstrates that the variable component has a virtually constant flux distribution, thereby confirming our assumption that this is the case.  If the spectrum did, in fact, change shape, the relationship in the flux-flux diagram would curve rather than be linear.  A lag between the continuum bands of a few days only introduces a small scatter about the straight line in the flux-flux diagrams, and our diagrams are consistent with this.

The slopes of the best-fit lines give the ratios of the variable component of the fluxes and hence the SED of the inner nucleus, as viewed from outside that galaxy.  Of course, the intrinsic SED of this region will differ due to extinction from dust in the outer parts of the nucleus.  Assuming some standard colours for the nuclear SED, the extinction can then be calculated by comparing the standard colours with the colours determined from the flux-flux diagrams.
Initially, we correct for the extinction in our Galaxy, $E(B-V)_g$, using values  from \citet*{schlegel98}.  Next, we correct for extinction within the nucleus of the AGN.  Assuming that the intrinsic SED for the nuclear region of an AGN follows $f_\nu \propto \nu^{1/3}$ in the optical, as expected for a classical thin accretion disc, we adjust the nuclear extinction, $E(B-V)_n$, so that the reddened model matches the observations.

This, of course, assumes that the difference between the spectra of the variable components and a $\nu^{1/3}$ spectrum is due to extinction.  As an independent check of the values we get from the flux variation gradient method, we also use the Balmer decrement method \citep[e.g.][]{reynolds97}.  As extinction by cosmic dust is highly wavelength dependent it can change observed line flux ratios significantly away from intrinsic values.  In this method, the observed H${\rm \alpha}$ to H${\rm \beta}$ broad line ratio is compared with an assumed intrinsic ratio, which we take to be 3.1 \citep{gaskell84} with the difference between the two determining the extinction required.  We use values in the literature for (H$\alpha$ +[NII])/H$\beta$ and H$\beta$/[OIII]$\lambda$5007, which are listed in a range of previous studies: \citet{anderson70,boksenberg75,cohen83,degrijp92,koski78,morris88,osterbrock76,osterbrock77,phillips78,rafanelli91,stephens89,winkler92}.  From these measured ratios we determine the H${\rm \alpha}$ to H${\rm \beta}$ broad line ratio assuming that the broad and narrow components are similarly reddened.  We do this by assuming that H$\beta$(narrow)/[OIII]$\lambda$5007 = 0.1 and H$\alpha$(narrow)/[NII] = 1, which are the approximate values where these ratios cluster for Seyfert 2 galaxies in the diagrams of \citet{veilleux87}.

\begin{figure}
  \centering
  \includegraphics[width=8cm]{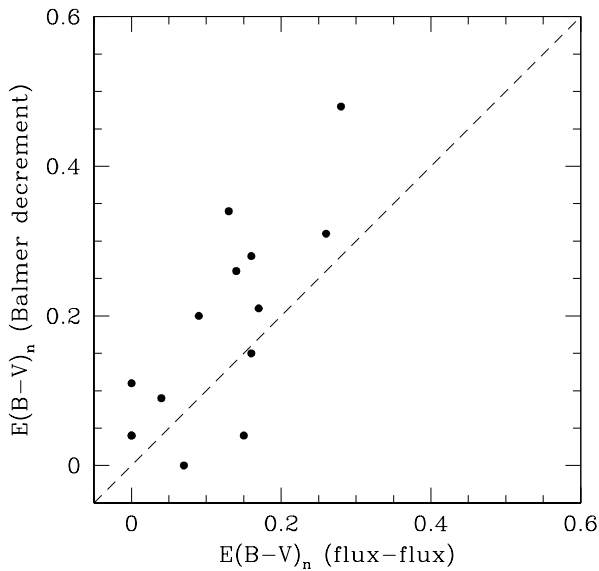}
  \caption[Comparison of $E(B-V)_n$ determined by the flux-flux and Balmer decrement methods]{Comparison of $E(B-V)_n$ determined by the flux-flux and Balmer decrement methods.  The dashed line indicates a one-to-one relationship.}
  \label{fig:ebmv}
\end{figure}

Using these two methods the nuclear extinctions have been calculated, and the results are presented in Table \ref{tab:ebv}.  A comparison of the reddening determined using these two methods is shown in Fig. \ref{fig:ebmv}.  Whilst the nuclear extinctions determined by these two methods do not agree exactly (largely due to observational difficulties in obtaining accurate spectrophotometric data by the authors of the various studies listed earlier), those AGN with higher nuclear extinction from one method also have high nuclear extinction from the other, and similarly for low values (see Fig. \ref{fig:ebmv}).  This confirms that there is significant extinction in the central regions of these AGN that needs to be accounted for in our models. 
\begin{table}
\centering
\caption[Nuclear extinction, $E(B-V)_n$, determined by the flux-flux and Balmer decrement methods]{Nuclear extinction, $E(B-V)_n$, determined by the flux-flux and Balmer decrement methods, as well as the extinction due to our own Galaxy, $E(B-V)_g$.  The total extinction is just the sum of both components.}
\label{tab:ebv}
  \begin{tabular}{cccccc}
    \hline
	Object & $E(B-V)_g$ & $E(B-V)_n$ & $E(B-V)_n$ \\
	 & & (flux-flux) & (Balmer decrement) \\
    \hline
NGC 4051  & 0.013 & 		0.17 & 0.21\\
NGC 4151  &  0.028 & 		0.15 & 0.04\\
NGC 3227 & 0.023 & 		0.26 & 0.31\\ 
NGC 3516  & 0.042 & 		0.16 & 0.15\\
NGC 7469  & 0.069 & 		0.04 & 0.09\\
NGC 5548  & 0.020 & 		0.16 & 0.28 \\
Mrk 6   & 0.136 & 		0.28 & 0.48\\
MCG+8-11-11 & 0.217 & 		0.13 & 0.34\\
Mrk 79  & 0.071 & 		0.09 & 0.20\\
Mrk 335 &  0.035 & 		0.07 & 0.00\\
Ark 120 & 0.128 & 		0.00 & 0.04\\
Mrk 509  & 0.057 & 		0.00 & 0.11\\
3C 390.3  & 0.071 & 		0.14 & 0.26\\
1E 0754.6+3928 & 0.066 &	0.00 & 0.04\\
    \hline
  \end{tabular}
\end{table}

\section{Results}

\begin{figure}
 \centering
 \includegraphics[width=8cm]{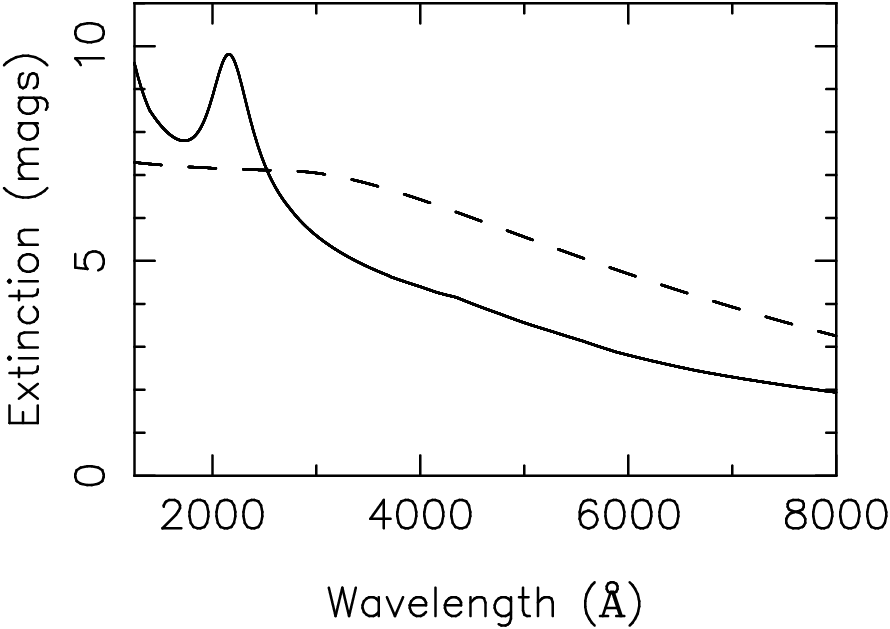}
 \caption{{\it Solid line}: the adopted extinction law of \citet{nandy75} and \citet{seaton79} evaluated at $E(B-V) = 1.0$ for R = 3.2. {\it Dashed line}: the AGN extinction law of \citet{gaskell04}, which has R = 5.15.  We have extended the \citet{gaskell04} law to longer wavelengths (determined up to $\lambda = 6250$\AA).  We use the \citet{nandy75} and \citet{seaton79} law to correct for the known reddening in our Galaxy, and the extinction law of \citet{gaskell04} to correct for the reddening intrinsic to the AGN.}
 \label{fig:extinction}
\end{figure}

We use the model described in the previous section to fit the data for all 14 AGN from the Sergeev et al. sample.  Specifically, we fit the model to the time delays, and the maximum and minimum SED by minimising $\chi^2$ using a downhill simplex \citep[the `amoeba' from Numerical Recipes in Fortran,][]{press92} to determine the best-fitting non-linear parameters, whilst optimising the linear parameters using optimal scaling.  For the time delays we adopt the peak of the CCF rather than the centroid.  The peak time delays are generally smaller than the centroid delays and therefore gives a lower measurement of the distance.  Firstly we correct for the known galactic extinction in the direction of each AGN using the combined galactic extinction law of \citet{nandy75} and \citet{seaton79}.  Further reddening due to extinction intrinsic to the AGN is included using the AGN reddening curve of \citet{gaskell04} (see Fig. \ref{fig:extinction}).  While these reddening laws both have very similar slopes in the optical, they differ greatly in the UV (where we are not concerned about here), and importantly, have different $R$ values, where $R\equiv A_V/E(B-V)$.

\begin{figure}
\centering
\includegraphics[angle=270,width=8cm]{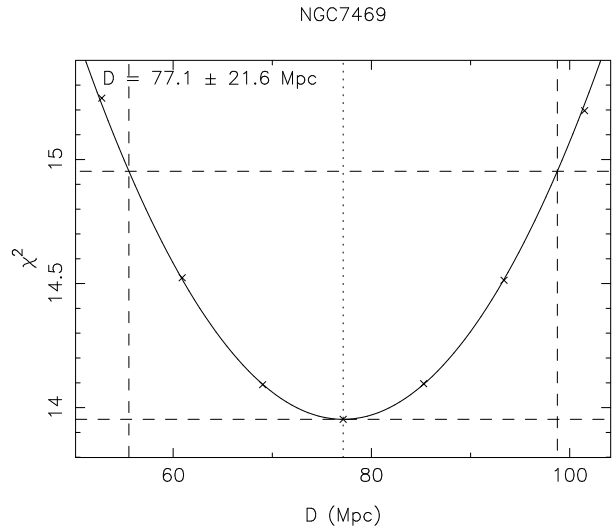}
\caption[$\chi^2$ parabola for the distance to NGC~7469]{$\chi^2$ error parabola for the distance to NGC~7469.  Solid line is the best-fit parabola to the $\chi^2$ values determined at a range of distances either side of the best-fit distance.  The dotted line indicates the location of the best-fit distance.  Dashed horizontal lines indicate $\chi^2_{min}$ and $\chi^2_{min} + 1$.  Dashed vertical lines indicate 1-$\sigma$ errors in the distance, i.e.,  the distances that correspond to $\chi^2_{min} + 1$.}
\label{fig:parabola}
\end{figure}

To determine the uncertainties in each of the fit parameters, we fix the parameter of interest at a value offset from the best-fit value.  The other parameters are then optimised and the $\chi^2$ value determined.  We do this for a range of values either side of the best-fit value, with the range chosen so that it covers $\Delta \chi^2 = 1.0$.  A parabola is then fit to the points, and forced to be a minimum (and hence have zero gradient) at the best-fit value.  From the parabola 1-$\sigma$ errors are determined from where $\Delta \chi^2 = 1.0$.  An example of this is shown in Fig. \ref{fig:parabola}.

We present the best-fitting models for each of the AGN in Fig. \ref{fig:modelfits}, and the best-fitting parameters in Table \ref{tab:modelfits}. Note that the extinction values are close to those determined by the two other methods (see Fig. \ref{fig:ebmv_all} for a comparison).  In several cases there are deviations of the difference spectrum from the best-fit model resulting in high values of reduced $\chi^2$.  This is likely to be due to contributions from broad emission lines.  For instance, H$\alpha$ lies in the R-band, and various Balmer lines and Fe II features lie in the B-band (see Fig. \ref{fig:3c390spec}). The R-band flux does sometimes appear higher than the model fit, particularly in cases where the $\chi^2$ is high, as would be expected due to contributions by H$\alpha$.

To check whether the determined galaxy spectra are realistic, we plot the galaxy spectra dereddened for galactic interstellar extinction, all normalised to the I band flux in Fig. \ref{fig:galspec}.  We do not deredden for the nuclear extinction that we have determined, as this reddening is expected to be due to dust in the nuclear region and not the galaxy as a whole. Additionally, we plot the galaxy spectrum for a Sab galaxy (as expected for a Seyfert) using the intrinsic galaxy colours of \citet{fukugita95}.  Whilst the majority of galaxy fluxes match the model galaxy in the V and R1 bands, all of the galaxies have a higher flux than the model in the R band.  We suggest that this is likely due to strong H$\alpha$ line emission from the AGN that falls within this band (see Fig. \ref{fig:3c390spec}), making the R band galaxy flux higher than for a typical Sab galaxy.  To check this we load the spectrum of 3C~390.3 shown in Fig. \ref{fig:3c390spec} into the synthetic photometry package {\sc Xcal}, and calculate the predicted flux in the R band.  We then subtract the H$\alpha$ line from this spectrum and calculate the predicted flux in the R band once again.  We find a factor of 1.3 difference between the 2 spectra.  Comparing the 3C~390.3 R band galaxy flux that we determine with that expected for an Sab galaxy we find a factor of 1.5 discrepancy.  Given that the Sab galaxy model is not going to perfectly match the actual galaxy spectrum of 3C 390.3, it seems reasonable that the observed difference could be due to H$\alpha$.  

There is one galaxy spectrum, 1E 0754.6+3928, that is significantly different from all the others.  In particular, the R1 flux is much higher than the other galaxies.  This object has the highest redshift of the sample, with $z = 0.096$.  The H$\alpha$ line will have therefore shifted from an emitted wavelength of 6563 $\mathrm \AA$ to 7193 $\mathrm \AA$, and into the R1 filter.  Note that the next highest redshift object in the sample is 3C 390.3, at $z = 0.056$, not high enough to shift the line significantly into the R1 band (see Fig. \ref{fig:3c390spec}).
\begin{figure*}
\centering
\includegraphics[width=16cm]{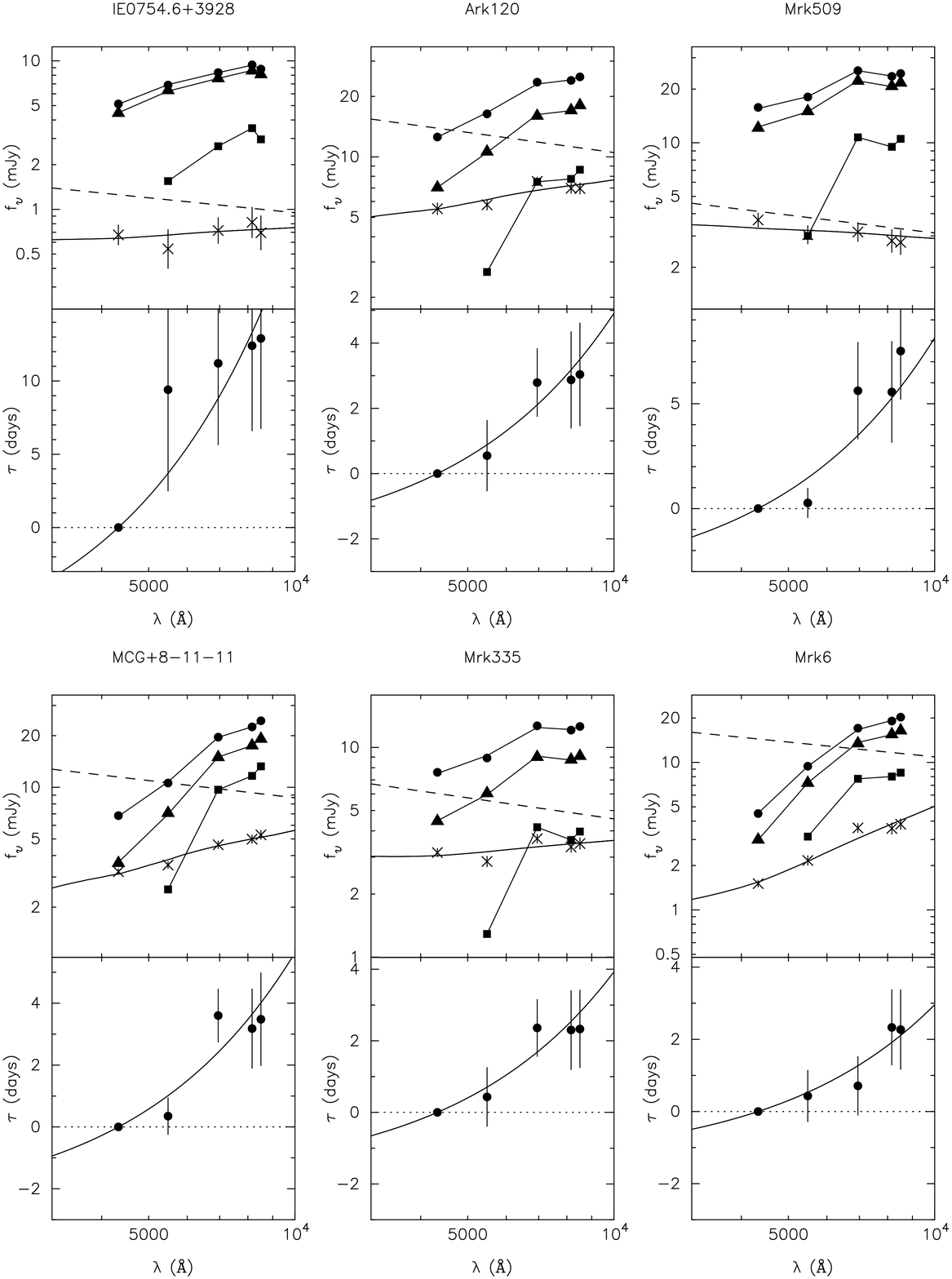}
\caption[Reprocessing model fits for all 14 AGN]{Reprocessing model fits for all 14 AGN. Each double panel (described next) shows the fit for one AGN. {\it Top:} Maximum (circles) and minimum (triangles) spectra, with best-fitting model.  Difference spectrum (crosses), with reddened accretion disc model.  De-reddened disk model is given by the dashed line.  Squares indicate best-fit galaxy fluxes.  {\it Bottom:} Time-delay distribution with best fit model.  Time delays relative to B-band, and determined from the peak of the CCF.  Figures continued on the next page.}
\label{fig:modelfits}
\end{figure*}
\setcounter{figure}{6}
\begin{figure*}
\centering
\includegraphics[width=16cm]{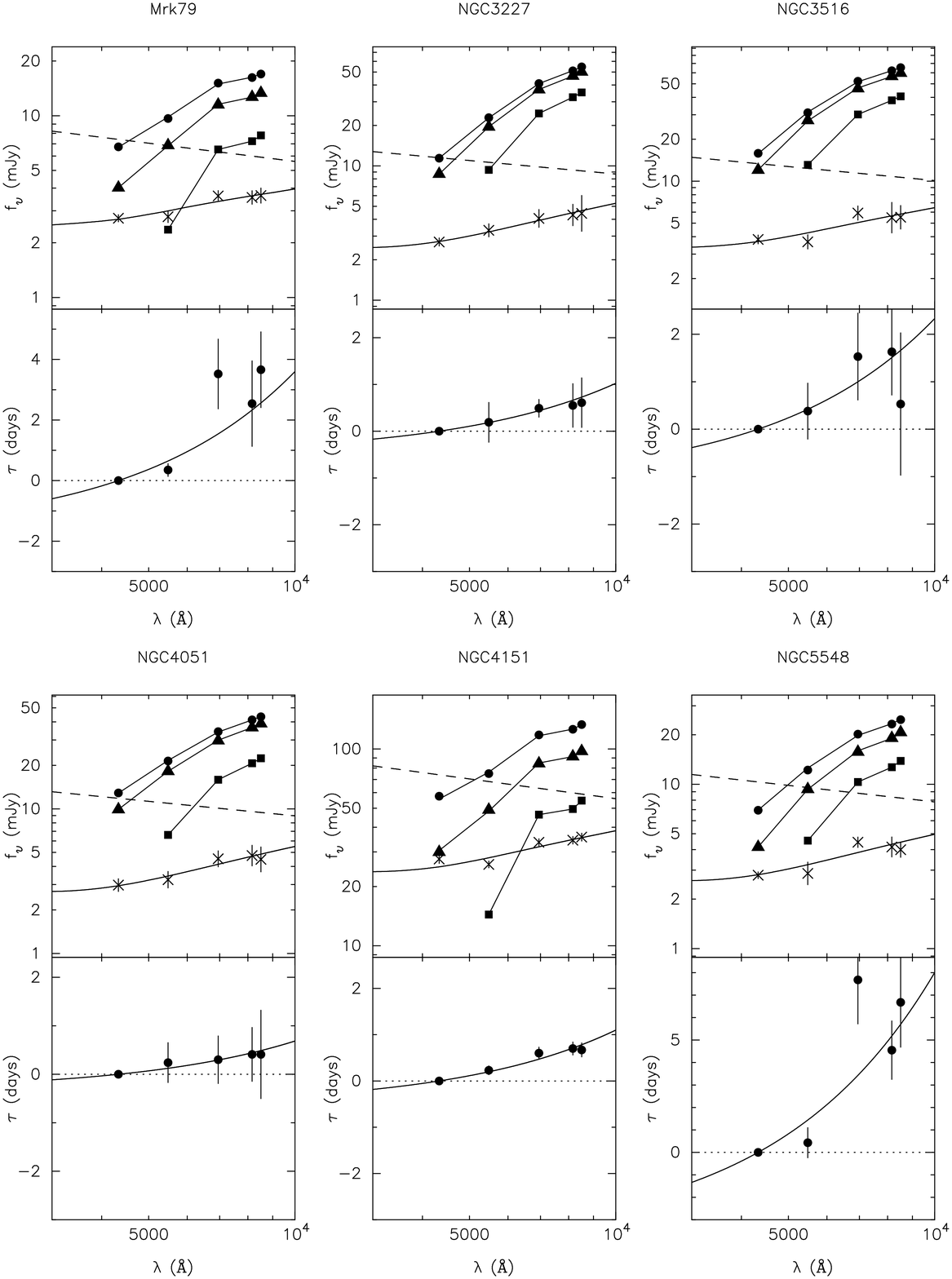}
\caption{continued.\hfil}
\end{figure*}
\setcounter{figure}{6}
\begin{figure*}
\centering
\includegraphics[width=10.66cm]{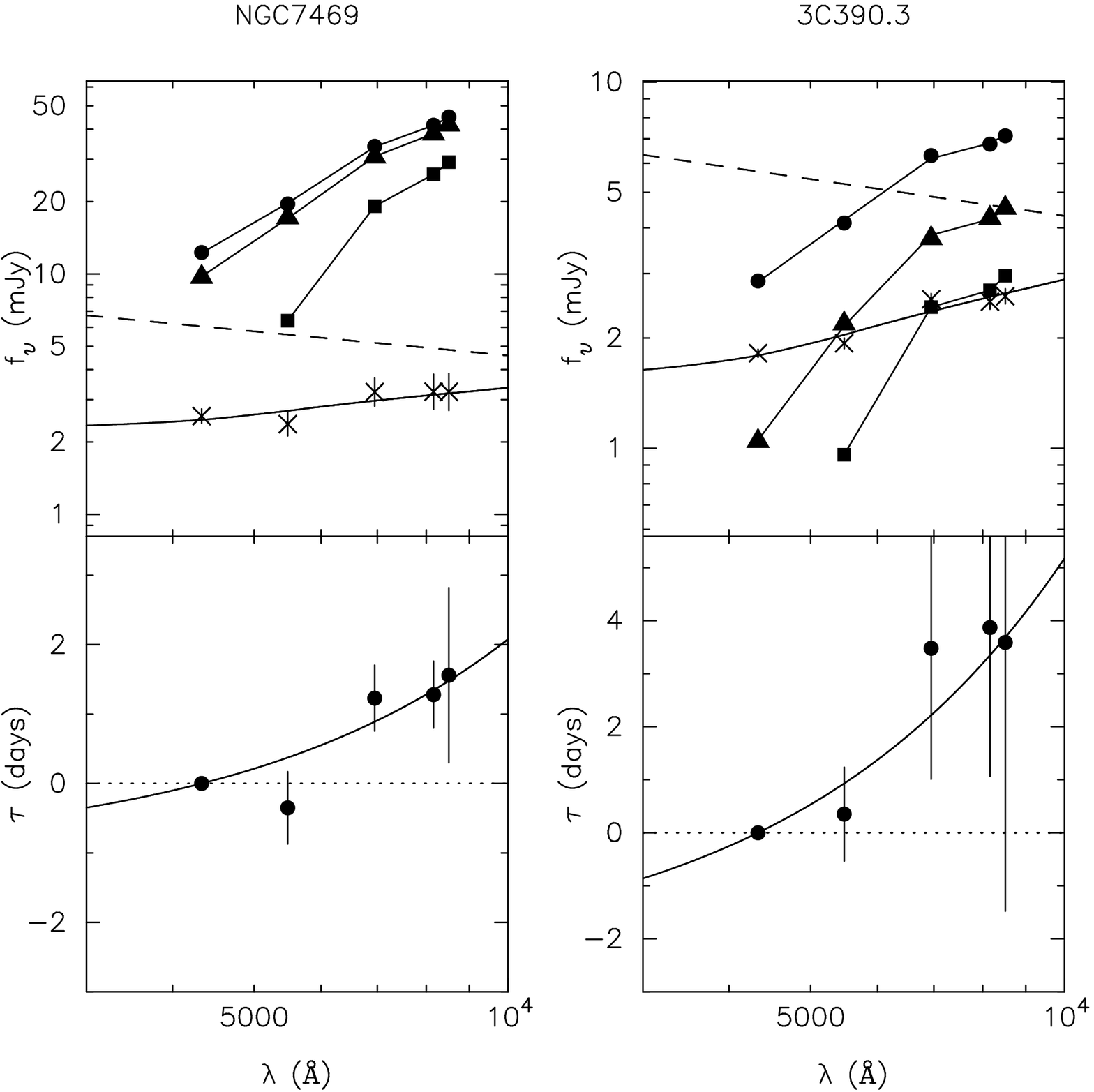}
\caption{continued.\hfil}
\end{figure*}

\begin{table*}
\centering
\caption[Best-fitting parameters from reprocessing model]{Best-fitting parameters from reprocessing model. $M \dot{M}$ is calculated using the faint state temperature. This gives a maximum value for this product, as the temperature in the faint state will not entirely be due to viscous heating and there will be some contribution from irradiation.}
\label{tab:modelfits}
  \begin{tabular}{ccccccccc}
    \hline
	Object & $z$ & $D$ & $cz/D$ & $E(B-V)_n$ & $T_F$ & $T_B$ & $M \dot{M}$ & $\chi^2_\nu$ \\
	& & (Mpc) & (km s$^{-1}$ Mpc$^{-1}$) & & ($10^3$ K) & ($10^3$ K) & ($10^7$ M$_\odot^2$ yr$^{-1}$) &\\
    \hline
	NGC4051       &   0.002    &   19 $\pm$  17 & 37 $\pm$ 33 &  0.24 $\pm$ 0.02 &  4.1 $\pm$  2.7 &  4.5 $\pm$  3.0 & 0.03 &  0.8 \\
NGC4151       &   0.003    &   19 $\pm$   2 & 52 $\pm$ 5  &  0.17 $\pm$ 0.01 &  5.4 $\pm$  0.5 &  6.8 $\pm$  0.6 & 0.09 &  8.1 \\
NGC3227       &   0.004    &   31 $\pm$  10 & 37 $\pm$ 12 &  0.24 $\pm$ 0.02 &  5.7 $\pm$  1.3 &  6.3 $\pm$  1.5 & 0.1  &  0.2 \\ 
NGC3516       &   0.009    &   63 $\pm$  27 & 42 $\pm$ 18 &  0.20 $\pm$ 0.02 & 10.3 $\pm$  3.3 & 11.4 $\pm$  3.6 & 1.1  &  3.5 \\
NGC7469       &   0.016    &   77 $\pm$  22 & 64 $\pm$ 18 &  0.11 $\pm$ 0.02 &  9.6 $\pm$  2.0 & 10.4 $\pm$  2.1 & 0.8  &  2.3 \\
NGC5548       &   0.017    &  341 $\pm$  62 & 15 $\pm$ 3  &  0.22 $\pm$ 0.01 & 24.6 $\pm$  3.3 & 29.9 $\pm$  4.0 & 36   &  5.9 \\
Mrk6          &   0.019    &   95 $\pm$  30 & 59 $\pm$ 19 &  0.30 $\pm$ 0.01 & 11.8 $\pm$  2.8 & 13.8 $\pm$  3.3 & 1.9  &  5.7 \\
MCG+8-11-11   &   0.020    &  248 $\pm$  49 & 25 $\pm$ 5  &  0.08 $\pm$ 0.01 & 18.4 $\pm$  2.7 & 23.3 $\pm$  3.4 & 12   &  3.9 \\
Mrk79         &   0.022    &  180 $\pm$  43 & 37 $\pm$ 9  &  0.13 $\pm$ 0.01 & 13.5 $\pm$  2.4 & 16.4 $\pm$  2.9 & 3.3  &  3.0 \\
Mrk335        &   0.026    &  219 $\pm$  53 & 35 $\pm$ 9  &  0.10 $\pm$ 0.01 & 14.4 $\pm$  2.6 & 17.5 $\pm$  3.2 & 4.2  &  7.3 \\
Ark120        &   0.032    &  190 $\pm$  51 & 51 $\pm$ 14 &  0.08 $\pm$ 0.01 & 16.8 $\pm$  3.3 & 20.9 $\pm$  4.1 & 7.9  &  5.7 \\
Mrk509        &   0.034    &  376 $\pm$  89 & 27 $\pm$ 6  &  0.00 $\pm$ 0.00 & 26.5 $\pm$  4.7 & 29.0 $\pm$  5.1 & 48   &  2.6 \\
3C390.3       &   0.056    &  400 $\pm$ 205 & 42 $\pm$ 22 &  0.16 $\pm$ 0.01 & 15.7 $\pm$  6.1 & 22.8 $\pm$  8.8 & 6.0  &  7.1 \\
IE0754.6+3928 &   0.096    & 1289 $\pm$ 130 & 22 $\pm$ 2  &  0.07 $\pm$ 0.03 & 54.3 $\pm$  9.4 & 57.0 $\pm$  4.6 & 856  &  1.0 \\

    \hline
  \end{tabular}
\end{table*}

\begin{figure}
  \centering
  \includegraphics[width=8cm]{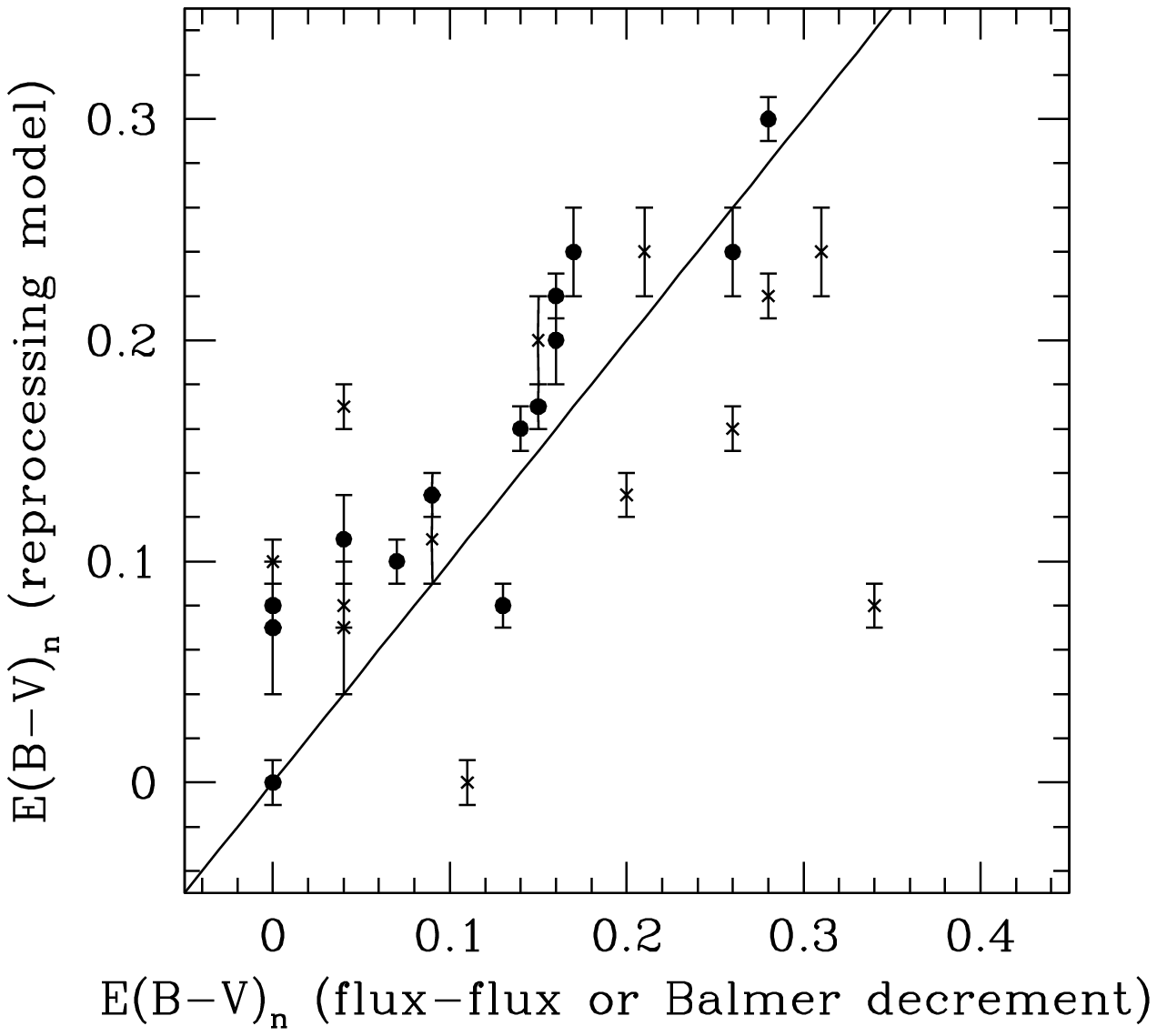}
  \caption[A comparison of the total E(B-V) values determined via the reprocessing model and the flux-flux and Balmer decrement methods]{A comparison of the total E(B-V) values determined via the reprocessing model and the flux-flux and Balmer decrement methods.  Circles indicate points determined by the flux-flux method, and crosses indicate those determined by the Balmer decrement method.  Dashed line indicates a one-to-one relationship.}
  \label{fig:ebmv_all}
\end{figure}

\begin{figure}
\centering
\includegraphics[width=8cm]{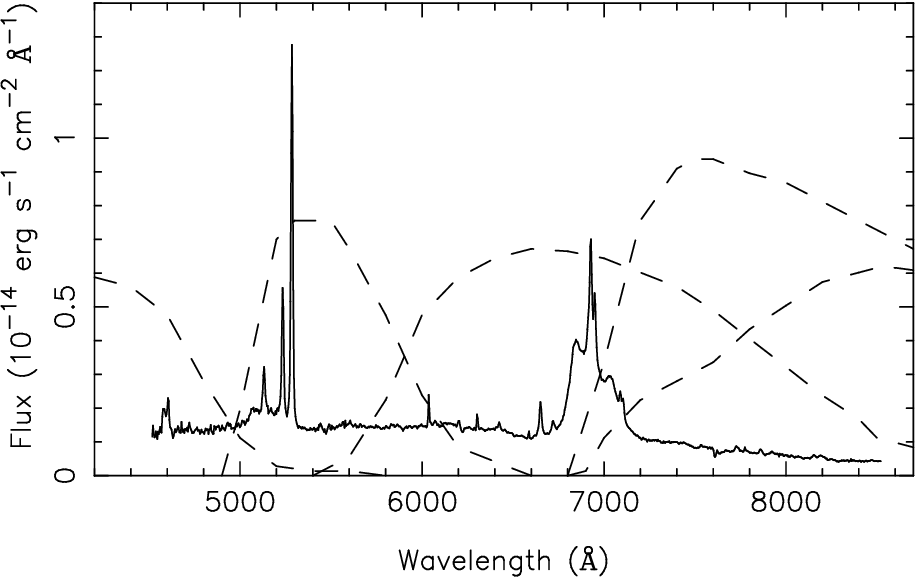}
\caption[Optical spectrum of 3C 390.3]{Optical spectrum of 3C 390.3, with the filter passbands shown (dashed line).  The filters (from left to right) are the Crimean B, V, R, R1 and I.  Note that H$\alpha$ sits in the R-band. 3C 390.3 is the object with the second highest redshift in our sample ($z = 0.056$).  Hence, in all other objects (expect 1E 0754.6+3928, which has $z = 0.096$), the H$\alpha$ line will be closer to its emitted wavelength of 6563 $\mathrm \AA$. Data from AGN Watch website (http://www.astronomy.ohio-state.edu/~agnwatch/).}
\label{fig:3c390spec}
\end{figure}

\begin{figure}
\centering
\includegraphics[width=8cm]{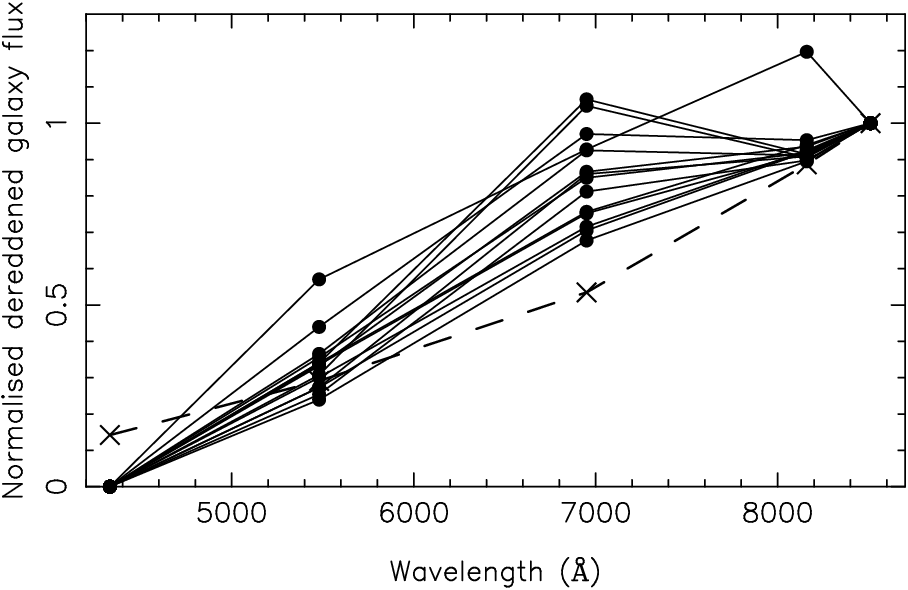}
\caption[Normalised dereddened galaxy spectra]{Dereddened galaxy spectra normalised to the I band (circles).  Note the B-band galaxy flux is forced to be zero in the fit.  The crosses show the galaxy spectrum for an Sab galaxy using the intrinsic galaxy colours of \citet{fukugita95}. }
\label{fig:galspec}
\end{figure}

\subsection{The Hubble Constant}

\begin{figure*}
\centering
\includegraphics[angle=270,width=17cm]{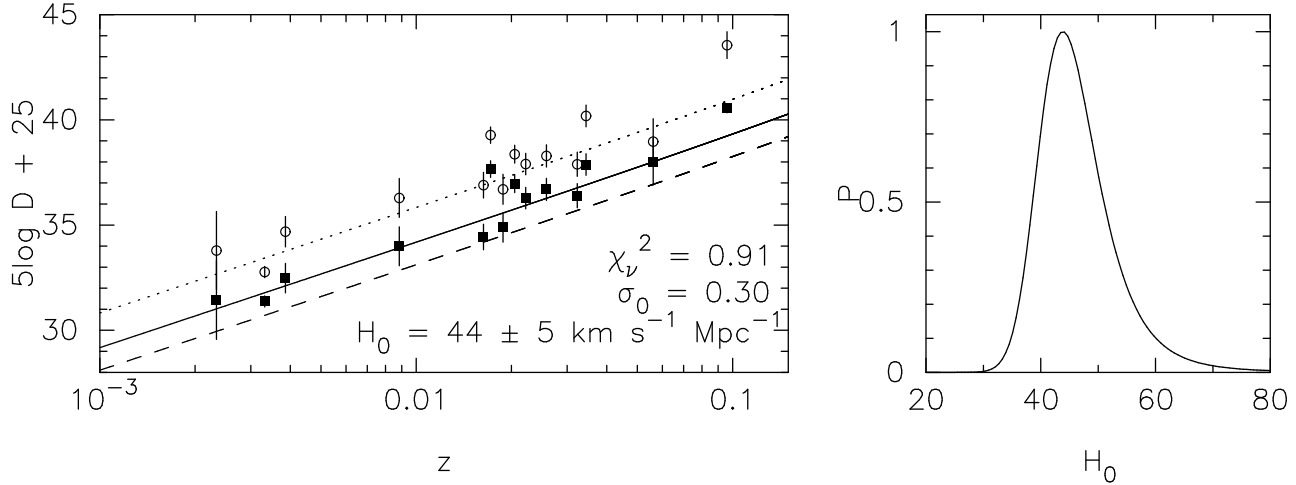}
\caption[Hubble diagram for 14 AGN]{{\it Left:} Hubble diagram for 14 AGN from the Sergeev et al. sample.  Distance modulus is plotted vs redshift. The solid line is the best-fitting model (fit to the filled circles), where as the dashed line indicates $H_0$ = 72 km s$^{-1}$ Mpc$^{-1}$. The filled circles are for the distances determined when the B-band galaxy flux is = 0, and the open circles are for when the B-band galaxy is set to the maximum possible.  The dotted line is the best-fit to the open circles.  {\it Right:} Probability distribution for $H_0$.}
\label{fig:H0}
\end{figure*}

Having measured distances to the 14 AGN using the continuum reverberation method detailed above, we are now in a position to determine Hubble's constant, $H_0$. Assuming a cosmological constant form for $\Omega_\Lambda$, in the Friedmann-Robertson-Walker metric, the luminosity distance is determined by:
\begin{eqnarray}
D_L &=& \frac{c}{H_0}(1+z)|\Omega_k|^{-1/2} S_k \bigg\{ |\Omega_k| ^{1/2} \times \nonumber \\
& &\int_1^{1+z} \frac{dx}{\left(\Omega_M x^3 + \Omega_\Lambda + \Omega_k x^2  \right)^{1/2}} \bigg\}
\label{eq:dl}
\end{eqnarray}
where $x=1+z$, $\Omega_k = 1 - \Omega_M - \Omega_\Lambda$ and $S_k$ is sinh for $\Omega_k > 0$ and sin for $\Omega_k < 0$.  For $\Omega_k = 0$, Eq. \ref{eq:dl} reverts to $cH_0^{-1}(1+z)$ times the integral \citep{carroll92}.  We assume the current standard cosmology with $\Omega_M = 0.3$ and $\Omega_\Lambda = 0.7$ when fitting this model.  As there is quite a large scatter in the distances (see Fig. \ref{fig:H0}), we also include a fractional systematic error, $f_0$, as a parameter in the fit to account for this intrinsic scatter. The uncertainty on a single distance measurement, $D$, is then given by $\sigma_i^2(D) = \sigma_D^2 + f_0^2 D^2$, where $\sigma_D^2$ is the measurement uncertainty on the distance from the fit. One cannot minimise $\chi^2$ when including such a term in the fit as continuing to increase $f_0$ will continue to decrease $\chi^2$ with no end. Instead we maximise the likelihood, which is equivalent to minimising $-2\ln L$, which includes a penalty for expanding the error bars:
\begin{equation}
-2\ln L = \chi^2 + 2\sum_{i=1}^N \ln(\sigma_i) + {\mathrm const} \; ,
\end{equation}
where there are $N$ data points, and $L\sim\exp(-\chi^2/2)/(\prod_i \sqrt{2\pi}\sigma_i)$.  The resulting fit is shown in Fig. \ref{fig:H0}.  We find, $H_0 = 44 \pm 5$ km s$^{-1}$ Mpc$^{-1}$ and $f_0 = 0.30$, with a reduced $\chi^2 = 0.91$.  The implications of this small value of $H_0$ are discussed in the next section.

To check the possible range in distance if the B-band galaxy flux is allowed to be non-zero (which it clearly will be), we also fit the disc reprocessing model when allowing the B-band galaxy flux to be the maximum possible value, i.e., equal to the minimum flux from the lightcurve.  Such a fit gives the maximum distance (and hence the minimum $H_0$) possible. Fitting to these distances gives $H_0 = 21 \pm 2$ km s$^{-1}$ Mpc$^{-1}$.  The distances and best-fitting model are shown in Fig. \ref{fig:H0}, as open circles and the dotted line, respectively.  This indicates that the possible range in $H_0$ if a reasonable B-band galaxy flux is determined for each galaxy will be between 21 and 44 km s$^{-1}$ Mpc$^{-1}$.  In any case, our conclusions still stand that too small a value of $H_0$ is determined by this model.

\section{Discussion}

Whilst the time-delay distributions are consistent with reprocessing in a disc, and fluxes and difference spectra can be fit by such a model (provided an estimate for the nuclear dust extinction is used), the distances determined using this method lead to a significantly smaller value of $H_0$ (at least a factor of $\sim$1.6) than determined by other methods - those that rely on the distance ladder e.g., the $HST$ Key Project  $H_0 = 72 \pm 8$ km s$^{-1}$ Mpc$^{-1}$ \citep{freedman01}, and other direct methods such as gravitational lensing time delays and the Sunyaev-Zeldovich effect \citep[e.g.][]{schmidt04,bonamente06,schechter05}.  All the observational evidence currently points towards a value of $H_0$ that is around 70 km s$^{-1}$ Mpc$^{-1}$.  As both direct methods and distance ladder methods both point to values of $H_0$ much greater than what we find, we interpret our result in terms of what this large difference in $H_0$ implies for the simple disc reprocessing model we have assumed.  We also note that there is a large scatter in the distances measured around the best-fit model.

We now briefly recap the main assumptions of our model, as this is relevant to our discussion.  Firstly, we assume that the variable ionizing radiation travels outwards at the speed of light, so that $R = c\tau$, and that there are no significant time delays other than the light travel time.  We also assume that the geometry is of a flat surface, and that there is azimuthal symmetry, so that the full annulus contributes, giving the surface area at radius $R$ as d(Area) $= 2\pi R \mathrm{d}R$.  We assume that there is thermal reprocessing, so that the relationship between surface brightness, wavelength and temperature is like that of a blackbody.  Finally, we correct the slope of the difference spectrum to $f_\nu \propto \nu^{1/3}$ assuming dust extinction from both our Galaxy and instrinsic to the AGN.  For the purposes of the following discussion, we list some scaling relations from our model.  The AGN flux $f \propto B \theta c^2 \tau^2 \cos i /D^2$, thus the distance to the object $D \propto c \tau (B \theta \cos i / f)^{1/2}$, and $H_0 \propto 1/D$.  Here $B$ is the surface brightness, and we assume $\theta = 2\pi$ (azimuthal symmetry).

\subsection{Scatter in distances}

In our Hubble diagram (Fig. \ref{fig:H0}), we note that there is a large scatter in the distances around the best-fitting cosmological model.  An extreme example of this is the two objects, NGC~7469 and NGC~5548, that have almost identical redshifts, yet, in the Hubble diagram they are separated by approximately 3 magnitudes in the distance modulus.  Whilst some of the scatter seen is likely due to adopting an inclination angle of $i = 45^{\circ}$ for all objects (as might be expected for a type I AGN), this effect is only small - the distance goes like $(\cos i)^{1/2}$.  According to unified AGN schemes \citep[e.g.][]{antonucci93}, for type I objects where we see the broad-emission line region, $i < 60^\circ$ and so $0.707 \leq (\cos i)^{1/2} \leq 1$.  Thus, whilst the inclination does cause scatter, it cannot account for the large difference between NGC~7469 and NGC~5548.  Comparing the lightcurves, ACF and CCFs for these two objects (Figs. \ref{fig:individ_lc}), it is seen that NGC~5548 is more slowly varying, and hence has a broader ACF, suggesting that maybe a difference between the two objects could be due to biased time delay measurements.

\begin{figure}
\centering
\includegraphics[width=6.45cm]{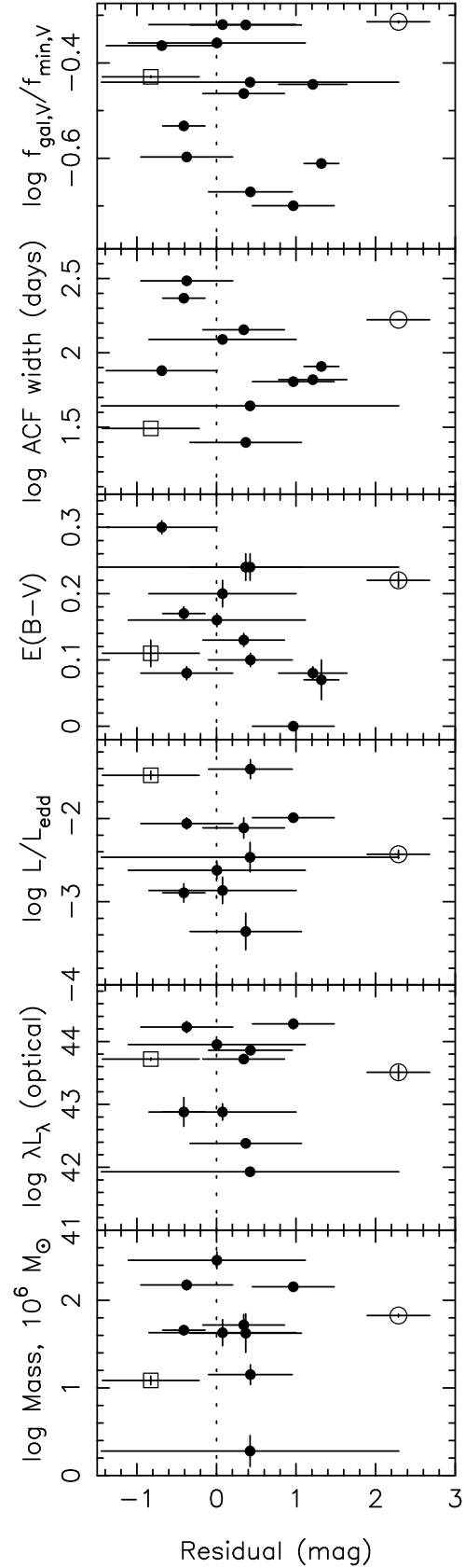}
\caption[Residuals compared to AGN properties]{The residuals of the Hubble diagram fit are compared with various AGN properties and parameters of the fit.  The open circle indicates NGC~5548, and the open square NGC~7469.  Here we use the reverberation mapping masses for these objects determined by \citet{petersonetal04}. }
\label{fig:AGN_residuals}
\end{figure}

To investigate this, and further possible causes of this scatter, Fig. \ref{fig:AGN_residuals} examines correlations between the residuals of the fit to the distances and various AGN properties and parameters of the fit.  Correlations with the mass, luminosity, fraction of Eddington luminosity, $E(B-V)$, ACF width, and fractional contribution of the galaxy to the minimum flux in the V were looked for, but importantly, none were found (see  Fig. \ref{fig:AGN_residuals}).  In this figure, we highlight NGC~7469 (open square) and NGC~5548 (open circle).  While the width of the ACFs differ largely between NGC~7469 and NGC~5548, there is no correlation between the ACF width and the residuals, as would be expected if this was the cause of the scatter.  Similarly, no correlation with $E(B-V)$ suggests that our adopted extinction law does not cause the scatter.
  
One possible explanation for the residuals could be due to the fact that we fix the galaxy in the B-band to be zero, as in our method we have no way of determining this.  This is clearly an incorrect assumption, and better constraints on the galaxy contribution may significantly improve matters.

Alternatively, reprocessing may not be the only source of the variability.  For instance, X-ray and optical lightcurves for NGC~5548 are seen to be correlated, but the fractional variability in the optical is larger than those seen in the X-rays \citep{uttley03}.  However, this is the long-term optical and X-ray variability, and the short-term variability is seen to be consistent with reprocessing in this object \citep{suganuma06}.

\subsection{Systematic offset in distance}

Having found that $H_0$ is at least a factor of 1.6 too small, we discuss possible causes for this systematic offset in distance.  One suggestion for the difference is that the lags measured via cross-correlation are biased to high values, hence giving high distances (and low $H_0$).  As the cross-correlation function (CCF) is the convolution of the delay map with the auto-correlation function (ACF), changes in the driving lightcurve properties (i.e. in the ACF), can lead to a different CCF, and hence a different lag, without changing the delay map.  If this is the case, one might expect to see a correlation between the residuals of the Hubble-diagram fit and the full-width half maximum of the ACF.  However, as discussed above, no such correlation is evident in Fig.~\ref{fig:AGN_residuals}.

Another possibility is that the irradiating source does not illuminate the entire disc.  If this is the case, then the solid angle for the reprocessed optical emission at each annulus in the disc, will be less than $2 \pi R \;{\mathrm d}R \; \cos i$ and hence the distances will be closer than determined and $H_0$ would be greater.  We briefly consider several possible causes of this.  Firstly, the irradiating source may be anisotropic, only illuminating specific parts of the disc (hence $\theta < 2\pi$).  Secondly, the surface of the disc may not be flat enough.  For instance, if the surface was bumpy, one can imagine that only the front side of the bump would be illuminated and that large bumps would actually place some of the disc in shadow.  For there to be a factor of $\sim 2$ increase in $H_0$, there would need to be a factor of 4 increase in the actual disc flux (as $f_\nu \propto D^{-2}$ and $H_0 \propto 1/D$).  Therefore one quarter of the disc would need to be illuminated, and the remaining three quarters not for the true disc flux to be four times larger than that observed.  In the lamppost model for disc illumination \citep[e.g.,][]{collin03} this would be hard to achieve as in this model the irradiating source is high above the central black hole and illuminates it in all directions.  However, if the incidence angle is low, so that irradiating photons can graze the disk surface, then small undulations in the disk surface, such as might arise from spiral density waves, may be sufficient to shadow a significant fraction of the disk surface.  Alternatively, if the illumination of the disc was more localised, for instance due to magnetic flares in the disc corona just above the disc surface \citep[e.g.,][]{collin03} then, it is easy to picture that only a small amount of the disc is irradiated.

It is important to consider whether additional extinction can be used to explain an increased true flux.  As extinction is wavelength-dependent, increasing its magnitude would also cause a change in shape of the spectral energy distribution.  Therefore only if the dust grains in the AGN are much larger than assumed (i.e.$R > 5.15$ that we used for the AGN extinction) could more extinction be allowed.

Furthermore, a scattering medium above the disc (such as a disc wind) could slow the outward propagation of the irradiating light, by making it more of a random walk diffusion process, rather than direct path to the disc surface.  this could increase the mean effective path length so that $\langle\tau\rangle > R/c$, and cause incident photons to be distributed over a wider solid angle than their direct line of sight.  Such a process would broaden the transfer function.  If scattering is important, then the detailed shape of the disk surface may be less important.

Finally, we also consider whether thermal reprocessing from dust can account for this offset in distance.  \citet{gaskell06} has recently proposed that the
short-wavelength tail of thermal emission from the dusty torus may account
for the observed increase of optical continuum time delay with wavelength.
However, the typical size of the dusty torus is thought to be on scales of up to hundreds of light days \citep{glass04,minezaki04,suganuma04,suganuma06}. The optical tail of the thermal emission from such dust would exhibit similar large delays and broadening, affecting broad structure in the cross-correlation function, but not significantly moving the cross-correlation peak that we are using here to measure inter-band delays.

Observationally, we can better understand AGN discs by increasing the cadence of multi-band monitoring to improve the delay measurements and ideally to measure delay distributions rather than just mean delays.  The main uncertainty in our distances arises from uncertainties in the delay measurements.  Defining the shape of the delay distribution, for comparison with predictions, would stringently test the blackbody disk model, and could detect or rule out additional significant broadening mechanisms.

\subsection{Possible cosmological probe?}

A systematic shift in the distance determination that is apparent in this analysis is not necessarily a problem when determining cosmological parameters other than $H_0$.  For $\Omega_M$ and $\Omega_\Lambda$, it is the {\it shape} of the redshift-distance relationship that matters.  It is therefore possible that even without understanding this offset, this method can be used as a cosmological probe provided that the systematic shift is redshift, and luminosity, independent.  As mentioned earlier, AGN are numerous throughout the Universe, out to high redshift.  As they are intrinsically extremely luminous objects, accurate photometry can be obtained out to redshifts far greater than supernovae type Ia can be observed, allowing similar accuracy in $\Omega_M$ and $\Omega_\Lambda$ with significantly more objects.  More luminous AGN will need to be observed at greater redshifts, however, and although more luminous quasars are known to be less variable, quasar variability increases toward the blue part of the spectrum \citep{vandenberk04}.  From studies of 25,000 quasars in the Sloan Digital Sky Survey, \citet{vandenberk04} also find a positive correlation of variability amplitude with redshift.  The redshift will move the observable region of the spectrum into the UV (in the emitted-frame), which is more variable, and likely has less galaxy contamination, however the time delays there will be shorter.  Counteracting this is time-dilation effects, which mean the observed time delay will be longer than the emitted time delay.

To explore possible constraints on $\Omega_M$ and $\Omega_\Lambda$ through monitoring programs, we simulate the distances to a sample of AGN. With the current dataset we have around 25\% rms errors in the distance, but the monitoring was only once every 2-3 days, improved temporal sampling should significantly improve the accuracy of the time-delay measured, and hence the distance determined.  We therefore assume that the distances can be measured to 15\% accuracy, which is plausible given high-cadence monitoring with a suite of robotic telescopes across the globe.  To approximate the redshift distribution of variable AGN that might be available in practice, we adopted a sample including the 14 nearby AGN in this work, augmented by 30 AGN that were discovered in the magnitude-limited QUEST1 variability survey \citep[Quasar Equatorial Survey Team, Phase 1,][]{rengstorf04}.  This provides known variable AGN over a range of redshifts.  In the simulation, we determine the distances from the redshift, assuming $\Omega_M = 0.3$, $\Omega_\Lambda = 0.7$ and $H_0 = 72$ km s$^{-1}$ Mpc$^{-1}$.  We add a random Gaussian-distributed error to each of the points, with a mean of 0.0, and 1-$\sigma$ = 15\%.  We then fit Eq \ref{eq:dl} by minimizing $\chi^2$.

\begin{figure*}
\centering
\includegraphics[angle=270,width=17cm]{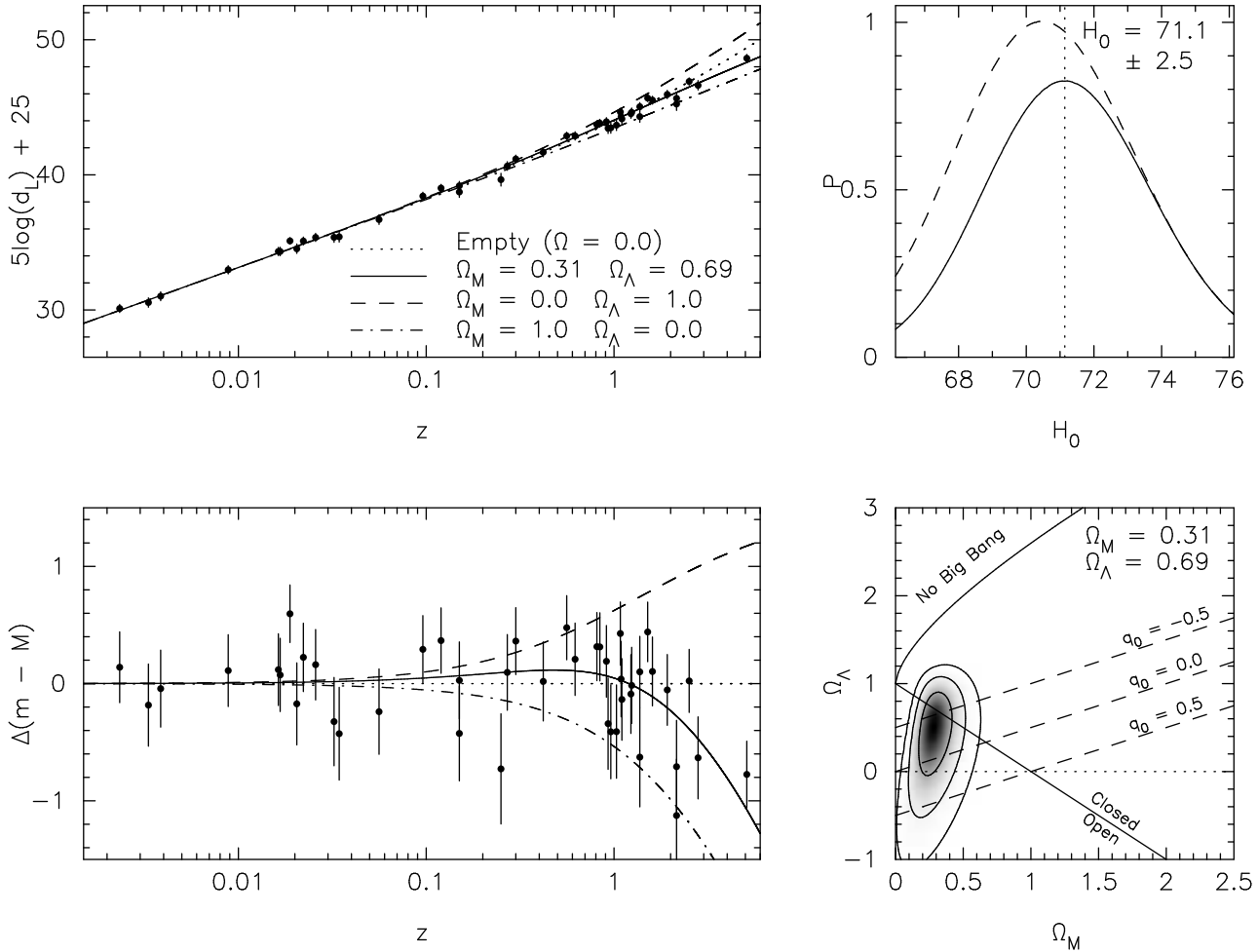}
\caption[Simulation of constraints on $H_0$, $\Omega_M$ and $\Omega_{\Lambda}$ from 44 AGN]{Simulation of constraints on $H_0$, $\Omega_M$ and $\Omega_{\Lambda}$ from 44 AGN.  {\it Top left:} Distance modulus vs. redshift for the 44 AGN.  Various cosmological models are shown, with the solid line indicating the best-fitting flat cosmology. {\it Bottom left:} Magnitude difference between the distance modulus and an empty Universe. {\it Top right:} Probability distribution for $H_0$.  The solid line indicates the probability distribution when a flat cosmology is assumed, and the dashed line shows the distribution with no constraint on $\Omega_{tot}$.{\it Bottom right:} Probability distribution for $\Omega_M$ and $\Omega_{\Lambda}$.  Contours indicate 1, 2 and 3-$\sigma$ confidence limits.}
\label{fig:AGNcosmo}
\end{figure*}

The results of this simulation are shown in Fig. \ref{fig:AGNcosmo}.  For comparison with our results, recent constraints from supernovae observations can be seen for example in Fig. 12 of \citet{woodvasey07}. For each value of $\Omega_M$ and $\Omega_\Lambda$ optimal scaling is used to determine the best value of $H_0$.  The best-fitting model for a flat Universe ($\Omega_{tot} = \Omega_M + \Omega_{\Lambda} = 1.0$) is indicated in each figure.  The probability distribution for $H_0$ in these figures is determined for a flat Universe and also when there is no constraint on $\Omega_{tot}$.  In the case of assuming a flat Universe, for each value of $H_0$ the best-fitting model and the probability are determined - the only free parameter is $\Omega_M$, as the condition that $\Omega_{tot} = 1.0$ determines $\Omega_{\Lambda}$.  Clearly, the input model is retrieved from the simulations, showing that this method has potential as a cosmological probe, if the systematics can be understood, or are redshift independent.

\section{Conclusions}

We have fitted the wavelength-dependent time delays and optical SEDs of 14 AGN using a disc reprocessing model, which has allowed a measure of the nuclear reddening in these AGN, as well as a measurement of the distances to the AGN.  However, the distances calculated using this method imply $H_0 = 44 \pm 5$ km s$^{-1}$ Mpc$^{-1}$, a factor of 1.6 less than the value that all other methods seem to be pointing to.  We have discussed the basic assumptions of the blackbody disc model and a number of possible systematic effets to which the simple blackbody disc model may be vulnerable.  Even with a systematic shift, it may be possible to use this method to probe $\Omega_M$ and $\Omega_{\Lambda}$, as it is the {\it shape} of the redshift-distance relationship that matters.  We have presented a simulation showing possible constraints from long-term monitoring of 44 AGN.

\subsection*{Acknowledgements}

EMC and HW gratefully acknowledge the support of PPARC.  The authors would like to thank Mike Goad, Ian M$\rm^{c}$Hardy, Simon Driver and Rick Hessman for stimulating discussions on this topic.   We also thank Luigi Gallo for helpful comments on a draft version of the paper.
 
\bibliographystyle{mn2e}
\bibliography{iau_journals,AGNdisks}

\end{document}